\documentclass[11pt,a4paper]{article}

\usepackage[T1]{fontenc}
\usepackage[utf8]{inputenc}
\usepackage{lmodern}
\usepackage{microtype}
\usepackage{amsmath,amssymb}
\usepackage{graphicx}
\usepackage{subcaption}
\usepackage{xcolor}
\usepackage{float}
\usepackage{hyperref}
\usepackage{mathrsfs}
\usepackage{cite} 
\usepackage{authblk} 

\hypersetup{
  colorlinks=true,
  linkcolor=blue,
  citecolor=blue,
  urlcolor=blue
}

\usepackage[margin=2.5cm]{geometry}

\usepackage{xcolor}
\usepackage[normalem]{ulem}
\newcommand{\MP}{M_{\rm P}}

\newcommand{\Hu}{H_u}
\newcommand{\Hd}{H_d}
\newcommand{\muhat}{\hat{\mu}}
\newcommand{\lamt}{\tilde{\lambda}}

\newcommand{\cw}{c_W}

\graphicspath{{figures/}}

\begin{document}



\def\un#1{\relax\ifmmode\@@underline#1\else
        $\@@underline{\hbox{#1}}$\relax\fi}


\let\under=\unt                 
\let\ced=\ce                    
\let\du=\du                     
\let\um=\Hu                     
\let\sll=\lp                    
\let\Sll=\Lp                    
\let\slo=\os                    
\let\Slo=\Os                    
\let\tie=\ta                    
\let\br=\ub                     


\def\a{\alpha}
\def\b{\beta}
\def\c{\chi}
\def\d{\delta}
\def\e{\epsilon}
\def\f{\phi}
\def\g{\gamma}
\def\h{\eta}
\def\i{\iota}
\def\j{\psi}
\def\k{\kappa}
\def\l{\lambda}
\def\m{\mu}
\def\n{\nu}
\def\o{\omega}
\def\p{\pi}
\def\q{\theta}
\def\r{\rho}
\def\s{\sigma}
\def\t{\tau}
\def\u{\upsilon}
\def\x{\xi}
\def\z{\zeta}
\def\D{\Delta}
\def\F{\Phi}
\def\G{\Gamma}
\def\J{\Psi}
\def\L{\Lambda}
\def\O{\Omega}
\def\P{\Pi}
\def\Q{\Theta}
\def\S{\Sigma}
\def\U{\Upsilon}
\def\X{\Xi}


\def\ve{\varepsilon}
\def\vf{\varphi}
\def\vr{\varrho}
\def\vs{\varsigma}
\def\vq{\vartheta}


\def\ca{{\cal A}}
\def\cb{{\cal B}}
\def\cc{{\cal C}}
\def\cd{{\cal D}}
\def\ce{{\cal E}}
\def\cf{{\cal F}}
\def\cg{{\cal G}}
\def\ch{{\cal H}}
\def\ci{{\cal I}}
\def\cj{{\cal J}}
\def\ck{{\cal K}}
\def\cl{{\cal L}}
\def\cm{{\cal M}}
\def\cn{{\cal N}}
\def\co{{\cal O}}
\def\cp{{\cal P}}
\def\cq{{\cal Q}}
\def\car{{\cal R}}
\def\cs{{\cal S}}
\def\ct{{\cal T}}
\def\cu{{\cal U}}
\def\cv{{\cal V}}
\def\cw{{\cal W}}
\def\cx{{\cal X}}
\def\cy{{\cal Y}}
\def\cz{{\cal Z}}


\def\Sc#1{{\hbox{\sc #1}}}      
\def\Sf#1{{\hbox{\sf #1}}}      



\def\slpa{\slash{\pa}}                            
\def\slin{\SLLash{\in}}                                   
\def\bo{{\raise-.3ex\hbox{\large$\Box$}}}               
\def\cbo{\Sc [}                                         
\def\pa{\partial}                                       
\def\de{\nabla}                                         
\def\dell{\bigtriangledown}                             
\def\su{\sum}                                           
\def\pr{\prod}                                          
\def\iff{\leftrightarrow}                               
\def\conj{{\hbox{\large *}}}                            
\def\ltap{\raisebox{-.4ex}{\rlap{$\sim$}} \raisebox{.4ex}{$<$}}   
\def\gtap{\raisebox{-.4ex}{\rlap{$\sim$}} \raisebox{.4ex}{$>$}}   
\def\TH{{\raise.2ex\hbox{$\displaystyle \bigodot$}\mskip-4.7mu \llap H \;}}
\def\face{{\raise.2ex\hbox{$\displaystyle \bigodot$}\mskip-2.2mu \llap {$\ddot
        \smile$}}}                                      
\def\dg{\sp\dagger}                                     
\def\ddg{\sp\ddagger}                                   

\font\tenex=cmex10 scaled 1200


\def\sp#1{{}^{#1}}                              
\def\sb#1{{}_{#1}}                              
\def\oldsl#1{\rlap/#1}                          
\def\slash#1{\rlap{\hbox{$\mskip 1 mu /$}}#1}      
\def\Slash#1{\rlap{\hbox{$\mskip 3 mu /$}}#1}      
\def\SLash#1{\rlap{\hbox{$\mskip 4.5 mu /$}}#1}    
\def\SLLash#1{\rlap{\hbox{$\mskip 6 mu /$}}#1}      
\def\PMMM#1{\rlap{\hbox{$\mskip 2 mu | $}}#1}   %
\def\PMM#1{\rlap{\hbox{$\mskip 4 mu ~ \mid $}}#1}       %
\def\Tilde#1{\widetilde{#1}}                    
\def\Hat#1{\widehat{#1}}                        
\def\Bar#1{\overline{#1}}                       
\def\sbar#1{\stackrel{*}{\Bar{#1}}}             
\def\bra#1{\left\langle #1\right|}              
\def\ket#1{\left| #1\right\rangle}              
\def\VEV#1{\left\langle #1\right\rangle}        
\def\abs#1{\left| #1\right|}                    
\def\leftrightarrowfill{$\mathsurround=0pt \mathord\leftarrow \mkern-6mu
        \cleaders\hbox{$\mkern-2mu \mathord- \mkern-2mu$}\hfill
        \mkern-6mu \mathord\rightarrow$}
\def\dvec#1{\vbox{\ialign{##\crcr
        \leftrightarrowfill\crcr\noalign{\kern-1pt\nointerlineskip}
        $\hfil\displaystyle{#1}\hfil$\crcr}}}           
\def\dt#1{{\buildrel {\hbox{\LARGE .}} \over {#1}}}     
\def\dtt#1{{\buildrel \bullet \over {#1}}}              
\def\der#1{{\pa \over \pa {#1}}}                
\def\fder#1{{\d \over \d {#1}}}                 


\def\frac#1#2{{\textstyle{#1\over\vphantom2\smash{\raise.20ex
        \hbox{$\scriptstyle{#2}$}}}}}                   
\def\half{\frac12}                                        
\def\sfrac#1#2{{\vphantom1\smash{\lower.5ex\hbox{\small$#1$}}\over
        \vphantom1\smash{\raise.4ex\hbox{\small$#2$}}}} 
\def\bfrac#1#2{{\vphantom1\smash{\lower.5ex\hbox{$#1$}}\over
        \vphantom1\smash{\raise.3ex\hbox{$#2$}}}}       
\def\afrac#1#2{{\vphantom1\smash{\lower.5ex\hbox{$#1$}}\over#2}}    
\def\partder#1#2{{\partial #1\over\partial #2}}   
\def\parvar#1#2{{\d #1\over \d #2}}               
\def\secder#1#2#3{{\partial^2 #1\over\partial #2 \partial #3}}  
\def\on#1#2{\mathop{\null#2}\limits^{#1}}               
\def\bvec#1{\on\leftarrow{#1}}                  
\def\oover#1{\on\circ{#1}}                              

\def\[{\lfloor{\hskip 0.35pt}\!\!\!\lceil}
\def\]{\rfloor{\hskip 0.35pt}\!\!\!\rceil}
\def\Lag{{\cal L}}
\def\du#1#2{_{#1}{}^{#2}}
\def\ud#1#2{^{#1}{}_{#2}}
\def\dud#1#2#3{_{#1}{}^{#2}{}_{#3}}
\def\udu#1#2#3{^{#1}{}_{#2}{}^{#3}}
\def\calD{{\cal D}}
\def\calM{{\cal M}}

\def\szet{{${\scriptstyle \b}$}}
\def\ulA{{\un A}}
\def\ulM{{\underline M}}
\def\cdm{{\Sc D}_{--}}
\def\cdp{{\Sc D}_{++}}
\def\vTheta{\check\Theta}
\def\fracm#1#2{\hbox{\large{${\frac{{#1}}{{#2}}}$}}}
\def\ha{{\fracmm12}}
\def\tr{{\rm tr}}
\def\Tr{{\rm Tr}}
\def\itrema{$\ddot{\scriptstyle 1}$}
\def\ula{{\underline a}} \def\ulb{{\underline b}} \def\ulc{{\underline c}}
\def\uld{{\underline d}} \def\ule{{\underline e}} \def\ulf{{\underline f}}
\def\ulg{{\underline g}}
\def\items#1{\\ \item{[#1]}}
\def\ul{\underline}
\def\un{\underline}
\def\fracmm#1#2{{{#1}\over{#2}}}
\def\footnotew#1{\footnote{\hsize=6.5in {#1}}}
\def\low#1{{\raise -3pt\hbox{${\hskip 0.75pt}\!_{#1}$}}}

\def\Dot#1{\buildrel{_{_{\hskip 0.01in}\bullet}}\over{#1}}
\def\dt#1{\Dot{#1}}

\def\DDot#1{\buildrel{_{_{\hskip 0.01in}\bullet\bullet}}\over{#1}}
\def\ddt#1{\DDot{#1}}

\def\DDDot#1{\buildrel{_{_{\hskip 0.01in}\bullet\bullet\bullet}}\over{#1}}
\def\dddt#1{\DDDot{#1}}

\def\DDDDot#1{\buildrel{_{_{\hskip 
0.01in}\bullet\bullet\bullet\bullet}}\over{#1}}
\def\ddddt#1{\DDDDot{#1}}

\def\Tilde#1{{\widetilde{#1}}\hskip 0.015in}
\def\Hat#1{\widehat{#1}}


\newskip\humongous \humongous=0pt plus 1000pt minus 1000pt
\def\caja{\mathsurround=0pt}
\def\eqalign#1{\,\vcenter{\openup2\jot \caja
        \ialign{\strut \hfil$\displaystyle{##}$&$
        \displaystyle{{}##}$\hfil\crcr#1\crcr}}\,}
\newif\ifdtup
\def\panorama{\global\dtuptrue \openup2\jot \caja
        \everycr{\noalign{\ifdtup \global\dtupfalse
        \vskip-\lineskiplimit \vskip\normallineskiplimit
        \else \penalty\interdisplaylinepenalty \fi}}}
\def\li#1{\panorama \tabskip=\humongous                         
        \halign to\displaywidth{\hfil$\displaystyle{##}$
        \tabskip=0pt&$\displaystyle{{}##}$\hfil
        \tabskip=\humongous&\llap{$##$}\tabskip=0pt
        \crcr#1\crcr}}
\def\eqalignnotwo#1{\panorama \tabskip=\humongous
        \halign to\displaywidth{\hfil$\displaystyle{##}$
        \tabskip=0pt&$\displaystyle{{}##}$
        \tabskip=0pt&$\displaystyle{{}##}$\hfil
        \tabskip=\humongous&\llap{$##$}\tabskip=0pt
        \crcr#1\crcr}}


\def\eV{\,{\rm eV}}
\def\keV{\,{\rm keV}}
\def\MeV{\,{\rm MeV}}
\def\GeV{\,{\rm GeV}}
\def\TeV{\,{\rm TeV}}
\def\sv{\left<\sigma v\right>}
\def\({\left(}
\def\){\right)}
\def\cm{{\,\rm cm}}
\def\K{{\,\rm K}}
\def\kpc{{\,\rm kpc}}
\def\beq{\begin{equation}}
\def\eeq{\end{equation}}
\def\bea{\begin{eqnarray}}
\def\eea{\end{eqnarray}}


\newcommand{\be}{\begin{equation}}
\newcommand{\ee}{\end{equation}}
\newcommand{\nbe}{\begin{equation*}}
\newcommand{\nee}{\end{equation*}}
\newcommand{\overbar}[1]{\mkern1.5mu\overline{\mkern-1.5mu#1\mkern-1.5mu}\mkern 1.5mu}
\newcommand{\fr}{\frac}
\newcommand{\lb}{\label}

\thispagestyle{empty}

\hbox to\hsize{%
\vbox{\noindent September 2026 \hfill IPMU26-0034 \hfill }}

\begin{center}
    \noindent {\Large \bfseries {Higgsino dark matter in the Starobinsky supergravity \\
    with the MSSM in light of the LUX-ZEPLIN event
}\par}
   
    \vspace{0.5cm}
   
    Daniel Frolovsky$^{a,b,*}$ and Sergei V. Ketov$^{c,d,\dagger}$
   
    \vspace{0.4cm}
   
    {\itshape \small
    $^a$ Institute for Theoretical Physics, Utrecht University,\\ Princetonplein 5, 3584 CC Utrecht, Netherlands \\[1.5mm]
    $^b$ Dutch Institute for Emergent Phenomena, Netherlands \\[1.5mm]
    $^c$ Department of Physics, Faculty of Science, Tokyo Metropolitan University, \\ 1-1 Minami-ohsawa, Tokyo 192-0397, Japan \\[1.5mm]
    $^d$ Kavli Institute for the Physics and Mathematics of the Universe (WPI),\\ The University of Tokyo Institutes for Advanced Study, Chiba 277-8583, Japan \\[1.5mm]
     \par
    }
\end{center}

\vspace{0.1cm}

{
\renewcommand{\thefootnote}{}
\footnotetext{$^\ast$ d.frolovskiy@uu.nl, the corresponding author}
\footnotetext{$^\dagger$ ketov@tmu.ac.jp}
}
\setcounter{footnote}{0}
\renewcommand{\thefootnote}{\arabic{footnote}}

\begin{abstract}
\noindent
We realize a nearly pure higgsino dark matter candidate with  mass of about $1$~TeV in the Minimal Supersymmetric Standard Model coupled to the Starobinsky supergravity. The recent LUX-ZEPLIN 248 keV nuclear-recoil event has renewed interest in higgsino dark matter. In this framework, the particle spectrum, including the Higgs boson mass, the higgsino mass and its splitting, is connected to cosmic inflation observables via gravitational mediation of supersymmetry breaking and subsequent renormalization group evolution. We show that the Higgs boson mass predicted by the model agrees with the measured value within the quoted uncertainties for a broad range of bino-induced mass splittings. Wino-induced splitting leads to a Higgs mass several GeV higher and is therefore excluded.
\end{abstract}

{\it Introduction $-$} In Ref.~\cite{Frolovsky:2026akv}, we proposed a framework connecting cosmic inflation to particle phenomenology by introducing the Starobinsky supergravity coupled to the Minimal Supersymmetric Standard Model (MSSM). The hidden sector responsible for inflation and spontaneous supersymmetry (SUSY) breaking emerges from the Einstein-frame description of Starobinsky supergravity. The cosmic microwave background (CMB) amplitude fixes the inflationary scale that sets the scale of the MSSM scalar soft terms via gravitational mediation of SUSY breaking. Requiring a light Higgs doublet and minimal stop mixing then fixes the vacuum and the boundary conditions for the renormalisation-group (RG) evolution of the Higgs quartic coupling. This relates the Higgs boson mass to the amplitude of primordial scalar perturbations. The gaugino masses depend on the gauge kinetic functions and can be constrained by dark matter (DM) phenomenology.

The particle nature of DM remains unknown. With conserved R-parity, the lightest supersymmetric particle (LSP) is stable and provides a DM candidate. In Ref.~\cite{Frolovsky:2026akv}, we considered two simple choices for the $\mu$ term: a "bare" mass term in the MSSM superpotential, and a term generated dynamically by interactions between the Higgs sector and the goldstino superfield of the Starobinsky supergravity. With any choice the higgsinos are heavy. Next, we identified a nearly pure wino LSP as a thermal DM candidate. The thermal relic abundance  selects a wino mass of about $3$~TeV. However, gamma-ray observations exclude such thermal wino as the sole DM component~\cite{Safdi:2025sfs}.

This motivates a combined $\mu$ term containing both "bare" and dynamical contributions in the same framework for the following reason. A tuned cancellation between these contributions allows the nearly pure higgsino with mass of about $1$~TeV to be the LSP, while preserving successful description of inflation. The thermal higgsino is  still a viable DM candidate ~\cite{Dessert:2022evk}.

An additional motivation comes from the $248$~keV nuclear-recoil event recently reported by the LUX-ZEPLIN Collaboration~\cite{LZ}. Although a single event does not establish a DM discovery, it has prompted studies of  possible inelastic higgsino interpretation and its constraints~\cite{FT,Yin,DiMauro,Pospelov:2026solar}.  In this Letter, we study higgsino DM in the framework~\cite{Frolovsky:2026akv}, relating it to cosmic inflation, spontaneous SUSY breaking and the observed Higgs boson mass. We test compatibility of the particle spectrum against the supergravity boundary conditions and the measured Higgs boson mass.

{\it Setup $-$} Starobinsky inflation \cite{Starobinsky:1980te} is a theoretically well-motivated approach for explaining the cosmological inflation, in very good agreement with Planck measurements of the cosmic microwave background (CMB) radiation \cite{BICEP:2021xfz}, see e.g., \cite{Ketov:2025nkr} for a modern review. It is possible to embed Starobinsky inflation into supergravity, both in the original formulation with the higher derivatives, known as the Jordan frame, and in the equivalent formulation in the Einstein frame without the higher derivatives 
\cite{Cecotti:1987sa,Gates:2009hu,Ketov:2010qz,Farakos:2013cqa,Ketov:2013dfa,Kehagias:2013mya,Addazi:2017rkc,Ellis:2018zya,Antoniadis:2024ypf}. In the standard 
Einstein frame, this leads to the appearance of two chiral superfields forming the hidden sector. Arranging an R-symmetry violation in the hidden sector \cite{Dalianis:2014aya} leads to spontaneous high-scale SUSY breaking with a Minkowski vacuum, whose scale and the corresponding masses of SUSY particles are related to the scale of inflation. In turn, this scale is fixed by the amplitude of primordial scalar perturbations. This supergravity model was dubbed the Starobinsky supergravity \cite{Frolovsky:2026akv}.

Starobinsky supergravity coupled to the MSSM in the Einstein frame has the following K\"ahler potential and superpotential  \cite{Frolovsky:2026akv}:
\begin{equation}
\mathcal{K}=-3\MP^2\ln\!\left[1+\fracmm{\mathcal{T}+\overline{\mathcal{T}}}{\MP}
+\gamma\,\fracmm{\mathcal{S}+\overline{\mathcal{S}}}{\MP}
-2\fracmm{\mathcal{S}\overline{\mathcal{S}}}{\MP^2}
+\fracmm{\zeta}{9}\fracmm{\mathcal{S}^2\overline{\mathcal{S}}^2}{\MP^4}\right]
+\sum_A\overline{\Phi}_A\Phi_A ,
\label{eq:K}
\end{equation}
\begin{equation}
\mathcal{W}=6m\,\mathcal{T}\mathcal{S}+\mathcal{W}_{\rm MSSM},\qquad
\mathcal{W}_{\rm MSSM}=\mu(\mathcal{S})\,\Hu\Hd+\sum_{A,B,C}y_{ABC}\Phi_A\Phi_B\Phi_C ,
\label{eq:W}
\end{equation}
where the two hidden-sector superfields $\mathcal{T}$ (inflaton) and $\mathcal{S}$ (goldstino)
arise from the dualisation of the Starobinsky supergravity, and $\Phi_A=(Q,L,d^c,u^c,e^c,\Hu,\Hd)$ are
the MSSM matter superfields. In Ref.~\cite{Frolovsky:2026akv}, the $\mu$ term was taken to be \emph{either} a constant
("bare") \emph{or} purely dynamical, $\mu_{\rm dyn}=\lamt\mathcal{S}^n/\MP^{n-1}$. Here we keep
both contributions and set $n=1$,
\begin{equation}
\mu(\mathcal{S})=\mu_0+\mu_{\rm dyn} .
\label{eq:mucomb}
\end{equation}
At the end of inflation the goldstino superfield acquires a non-zero vacuum expectation value,
$\VEV{\mathcal{S}}=s_0\MP$, and SUSY is spontaneously broken.
The Minkowski vacuum conditions fix
$\gamma=2(1-s_0^2)/(3s_0)$ and $\zeta=(1+2s_0^2)/(3s_0^4)$, while  gravitino acquires the mass ~\cite{Frolovsky:2026akv}
\begin{equation} \lb{gravitinom}
m_{3/2}^2=\fracmm{1}{\MP^4}\VEV{e^{\mathcal{K}/\MP^2}\abs{\mathcal{W}}^2}
=\dfrac{2187\,m^2 s_0^2 (1+2s_0^2)^2}{8(11-5s_0^2)^3}~,
\end{equation}
where $m$ is the inflationary mass scale. We use the observed scalar amplitude
$A_s\approx2.1\times10^{-9}$ 
at the pivot scale $k_*=0.05\,{\rm Mpc}^{-1}$ and take $N_*=55$ e-folds before the end of inflation.
Given $s_0\approx\sqrt{13/28}$, as is obtained below, the normalization of the scalar power spectrum yields
$m\approx1.80\times10^{13}$~GeV. Equation~(\ref{gravitinom}) then leads to
$m_{3/2}\approx1.53\times10^{13}$~GeV.

In the flat limit $\MP\to\infty$ at fixed $m_{3/2}$
 the soft scalar masses, trilinear and bilinear couplings
can be derived \cite{Brignole:1997wnc,Brignole:2010sax}. With the canonical matter K\"ahler metric, the soft scalar masses are universal,
\begin{equation}
m_{\bar{\alpha}\beta}^{2}=m_{3/2}^2\,\delta_{\bar{\alpha}\beta}
\quad\mbox{or}\quad m_0=m_{3/2}~,
\end{equation}
while the universal trilinear parameter is given by \cite{Frolovsky:2026akv}
\begin{equation}
A=m_{3/2}\left[3-\fracmm{2(11-5s_0^2)}{3(1+2s_0^2)}\right]
=m_{3/2}\,\fracmm{28s_0^2-13}{3(1+2s_0^2)}~.
\end{equation}
  The physical MSSM $\mu$ parameter is obtained after supergravity rescaling
\begin{equation}
\muhat=e^{\hat{\mathcal{K}}/2}\,\mu .
\label{eq:muhat}
\end{equation}
The bilinear soft parameter in our case is 
\begin{equation}
B=A-m_{3/2}+F^m\partial_m\ln\mu
=A-m_{3/2}+\fracmm{\mu_{\rm dyn}}{\mu}\,\fracmm{4(11-5s_0^2)}{9(1+2s_0^2)}\,m_{3/2} .
\label{eq:Bmu}
\end{equation}

Gaugino masses are determined by the holomorphic gauge kinetic function $f_a$,
\begin{equation}
M_a=\fracmm{1}{2}\,({\rm Re}f_a)^{-1}F^m\partial_m f_a ,
\qquad {\rm Re}f_a=\fracmm{1}{g_a^2} .
\label{eq:Ma}
\end{equation}
After taking the ansatz $f_a=c_1^{(a)}+c_S^{(a)}\mathcal{S}/\MP$ we obtain
\begin{equation}
M_a^{\rm tree}=\fracmm{1}{2}\,g_a^2\,c_S^{(a)}\,m_{3/2}\,s_0\,
\fracmm{4(11-5s_0^2)}{9(1+2s_0^2)} ,
\qquad
c_1^{(a)}+c_S^{(a)}s_0=\fracmm{1}{g_a^2} .
\end{equation}
 The total gaugino mass at the matching scale is a sum of the
tree-level and anomaly-mediated terms, $M_a^{\rm total}=M_a^{\rm tree}+M_a^{\rm anom}$, with
\begin{equation} \label{AMSB_masses}
M_1^{\text{anom}}=\fracmm{33}{5}\fracmm{\alpha_1}{4\pi}(m_{3/2}-A)~,\quad
M_2^{\text{anom}}=\fracmm{\alpha_2}{4\pi}(m_{3/2}-5A)~,\quad
M_3^{\text{anom}}=-\fracmm{3\alpha_3}{4\pi}(m_{3/2}+A)~.
\end{equation}

Demanding an eigenvalue of the tree-level Higgs mass-squared matrix be well below the
SUSY-breaking scale implies 
\begin{equation}
\det\mathcal{M}_H^2=(m^2_{H_u}+\muhat^2)(m^2_{H_d}+\muhat^2)-(B\muhat)^2=0 .
\label{eq:det}
\end{equation}
With the universal scalar masses, $m^2_{H_u}=m^2_{H_d}=m_{3/2}^2$, it reads
\begin{equation}
m_{3/2}^2+\muhat^2=\left|B\muhat\right|.
\label{eq:det2}
\end{equation}
Inserting this to the tree-level minimization condition gives
\begin{equation}
\sin2\beta=\fracmm{2B\muhat}{m^2_{H_u}+m^2_{H_d}+2\muhat^2}
=\fracmm{B\muhat}{m_{3/2}^2+\muhat^2}=1
\quad\ \rm{and,~ therefore,}   \quad \tan\beta=1.
\label{eq:tanb}
\end{equation}

 		At the matching scale $m_0$, the MSSM higgsino mass parameter is set by the tuned cancellation between the bare and dynamical terms,
\begin{equation}
\muhat(m_0^+)=e^{\hat{\mathcal{K}}/2}\left(\mu_0+\lamt s_0\MP\right) .
\label{eq:mu_boundary}
\end{equation}
Here $m_0^+$ ($m_0^-$) denotes the value before (after) integrating out the heavy fields at the same scale $m_0$.
Imposing the vanishing stop mixing, $X_t=A_t-\muhat(m_0^+)\cot\beta=0$, amounts to $A\approx\muhat(m_0^+)$ that
fixes
\begin{equation}
s_0=\sqrt{13/28} 
\end{equation}
in the leading order with respect to $\muhat/m_{3/2}$. Therefore, we get $m_0=m_{3/2}\approx1.5\times10^{13}$~GeV.

{\it Dark matter $-$}  Higgsino of mass 1~TeV with  neutral mass splitting $\delta\approx350$~keV  was offered as a
possible interpretation of the $248\pm23\,({\rm stat})\pm23\,({\rm sys})$~keV nuclear-recoil event
reported by LUX-ZEPLIN collaboration~\cite{LZ}. This interpretation was discussed both in the LZ analysis
(the inelastic $\mathcal{O}_1^s$ model) and in the dedicated studies~\cite{FT,Yin,DiMauro}. However, as was shown in~\cite{Pospelov:2026solar}, the non-observation of high-energy neutrinos from the Sun by the IceCube experiment~\cite{IceCube:2025fcu} leads to the robust limit $\delta>566$~keV for thermal higgsinos constituting all DM,  which excludes the higgsino interpretation of the LUX-ZEPLIN event.

Pure higgsino is a  degenerate Dirac state. The splitting is generated by mixing with 
gauginos and integrating them out with 
\begin{equation}
\delta=m_{\chi_2^0}-m_{\chi_1^0}\approx
m_Z^2\left(\fracmm{\sin^2\theta_W}{M_1}+\fracmm{\cos^2\theta_W}{M_2}\right) \,,
\label{eq:delta}
\end{equation}
where $\theta_W$ is the Weinberg angle. We consider two cases, where the splitting is
realized either by $M_1$ with $M_2=m_0$ or by $M_2$ with $M_1=m_0$. Including threshold matching and RG evolution, we find the running gaugino masses at the respective thresholds for the benchmark $\delta=350$~keV as
\begin{align}
&M_1\approx6.68\times10^{6}\ \mbox{GeV}
\qquad(M_2=m_0), \nonumber\\
&M_2\approx1.65\times10^{7}\ \mbox{GeV}
\qquad(M_1=m_0).
\label{eq:M1M2}
\end{align}
The splitting $\delta$ and the thermal relic condition constrain pole masses at low energies, whereas (\ref{gravitinom}--\ref{eq:mu_boundary}) are the boundary conditions at the scale $m_0$.
The two sets are therefore not directly comparable, and the low-scale inputs must first be evolved
upwards through the effective field theory (EFT) tower before the framework can be tested. We do this by
evolving the higgsino and gaugino mass parameters to $m_0$, matching them to the supergravity conditions, and then running back down to obtain the Higgs boson mass. The resulting  high-scale boundary conditions are
\begin{align}
&m_{H_u}=m_{H_d}=m_0=m_{3/2}\approx1.53\times10^{13}\ \mbox{GeV},\qquad
\tan\beta\approx1.01, \nonumber\\
&s_0\approx0.681697,\quad M_1(m_0^-)\approx6.80\times10^{6}\ \mbox{GeV},\quad M_2=M_3=m_0, \nonumber\\
&A\approx\muhat(m_0^+)\approx3.2\times10^{10}\ \mbox{GeV},\quad \muhat(m_0^-)\approx5.49\times10^{4}\ \mbox{GeV}, \nonumber\\
&\mbox{and}\quad s_0\approx0.681446,\quad
M_2(m_0^-)\approx1.18\times10^{7}\ \mbox{GeV},\quad M_1=M_3=m_0, \nonumber\\
&A\approx\muhat(m_0^+)\approx6.2\times10^{9}\ \mbox{GeV},\quad \muhat(m_0^-)\approx7.09\times10^{5}\ \mbox{GeV}.
\label{eq:bc}
\end{align}
The two higgsino mass parameters obey $\muhat(m_0^-)=\muhat(m_0^+)+\Delta\mu$, where $\Delta\mu$ is the finite threshold correction from the heavy fields. The cancellation in Eq.~(\ref{eq:mu_boundary}) is tuned to give the required light higgsino mass after this correction and RG evolution.

\begin{table}[h]
\centering
\begin{minipage}{0.47\textwidth}\centering
\small  light bino\\[2pt]
\begin{tabular}{cccc}
\hline
$a$ & $g_a(m_0^+)$ & $c_1^{(a)}$ & $c_S^{(a)}$ \\
\hline
$1$ & $0.5567$ & $3.26856$ & $-0.06122$ \\
$2$ & $0.5617$ & $0.00410$ & $4.64348$ \\
$3$ & $0.5776$ & $-0.02111$ & $4.42845$ \\
\hline
\end{tabular}
\end{minipage}\hfill
\begin{minipage}{0.47\textwidth}\centering
\small light wino \\[2pt]
\begin{tabular}{cccc}
\hline
$a$ & $g_a(m_0^+)$ & $c_1^{(a)}$ & $c_S^{(a)}$ \\
\hline
$1$ & $0.5567$ & $0.04135$ & $4.67471$ \\
$2$ & $0.5842$ & $2.93659$ & $-0.00927$ \\
$3$ & $0.5776$ & $-0.01941$ & $4.42739$ \\
\hline
\end{tabular}
\end{minipage}
\caption{MSSM gauge couplings and gauge kinetic function coefficients from the leading reconstruction at $m_0^+$.}
\label{tab:gauge}
\end{table}

The EFT towers below $m_0$ are 
\begin{align}
\text{MSSM}&\ \xrightarrow{\ m_0\ }\ \text{SM}+\widetilde{H}+\widetilde{B}
\ \xrightarrow{\ M_1\ }\ \text{SM}+\widetilde{H}
\ \xrightarrow{\ \mu\ }\ \text{SM}, \nonumber\\
\text{MSSM}&\ \xrightarrow{\ m_0\ }\ \text{SM}+\widetilde{H}+\widetilde{W}
\ \xrightarrow{\ M_2\ }\ \text{SM}+\widetilde{H}
\ \xrightarrow{\ \mu\ }\ \text{SM},
\label{eq:tower}
\end{align}
in the light-bino and light-wino cases, respectively. The two remaining gauginos are
integrated out at $m_0$ together with the scalars and the heavy Higgs fields.
The RG evolution and Higgs mass calculation are performed using SARAH~\cite{Staub:2013tta} and FlexibleSUSY~\cite{Athron:2014yba,Athron:2017fvs}, with SplitMSSM adapted to the EFT towers in (\ref{eq:tower}). We use two-loop RG evolution in the intermediate EFTs and one-loop matching of renormalizable parameters, supplemented by selected two-loop threshold corrections to the Higgs quartic coupling~\cite{Bagnaschi:2017xid}. The SM running includes selected three- and four-loop contributions, while the Higgs pole mass includes selected two- and three-loop terms and the leading four-loop QCD correction. We include one-loop matching of the neutral-splitting operators at the light-gaugino threshold using Matchete~\cite{FuentesMartin:2022matchete}, their two-loop evolution with anomalous dimensions adapted from \cite{Ibarra:2024weinberg}, and the one-loop pole-splitting correction.

\begin{figure}[t]
\centering
\includegraphics[width=0.5\textwidth]{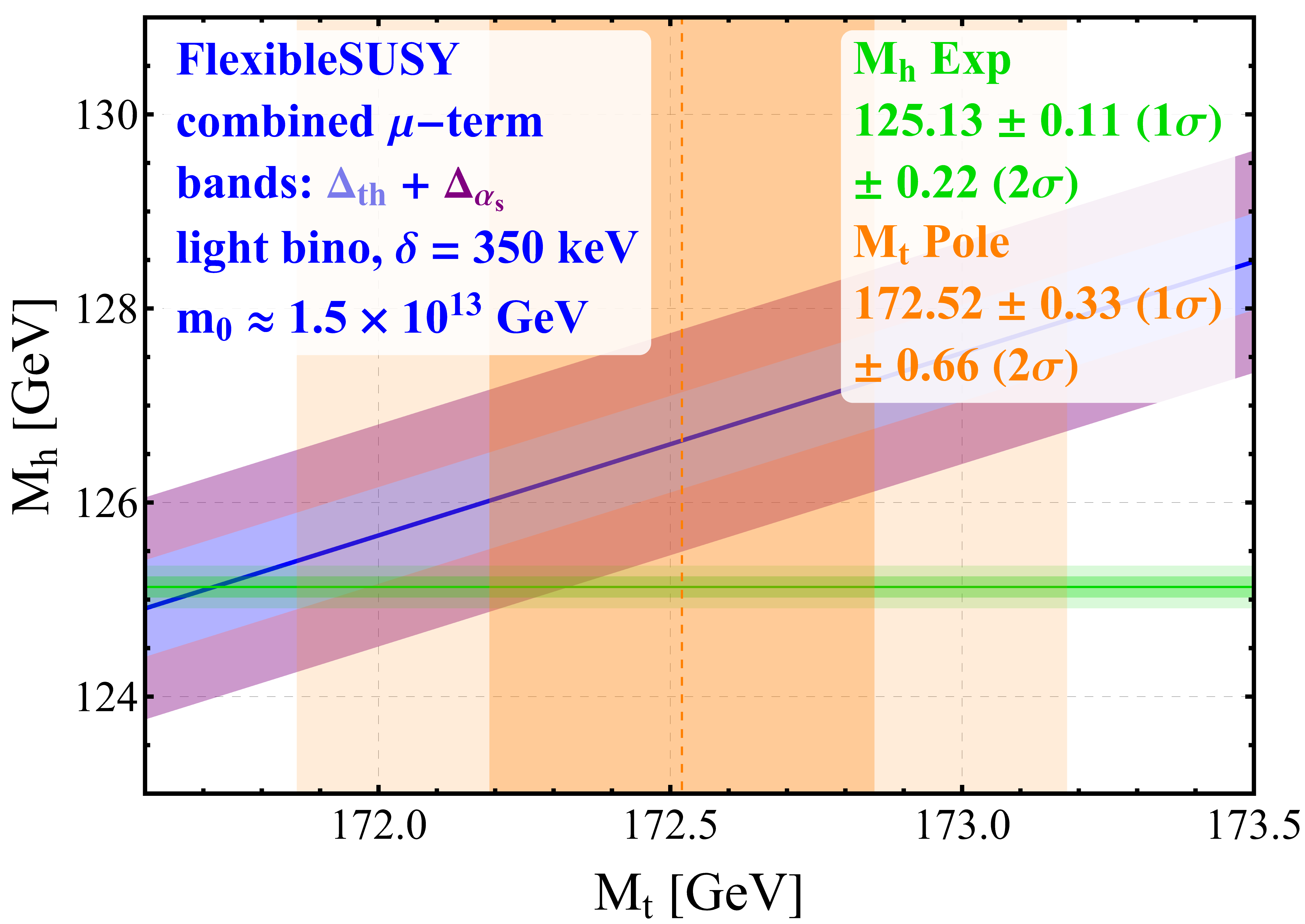}\hfill
\includegraphics[width=0.5\textwidth]{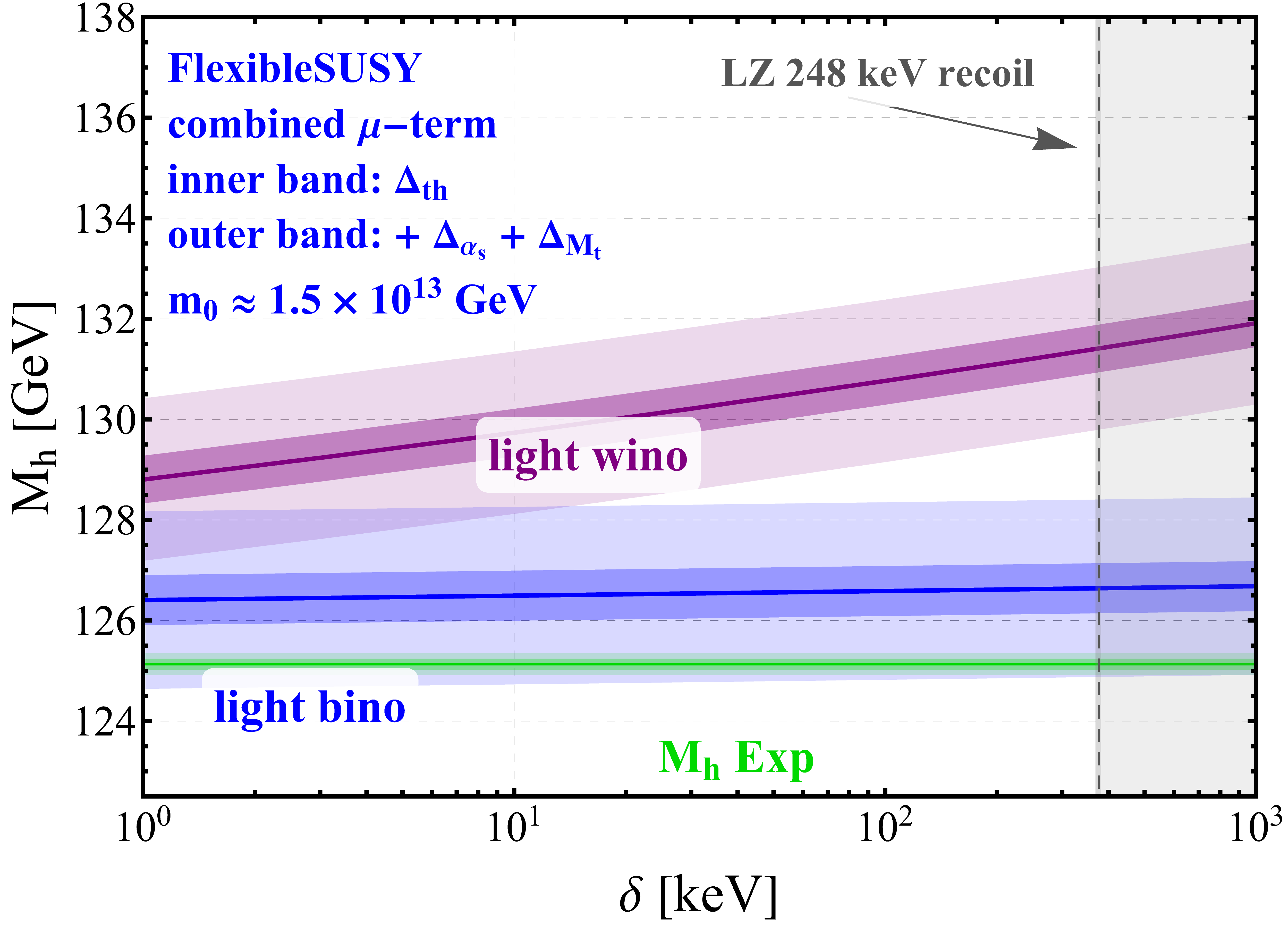}
\caption{Higgs boson mass against the top quark pole mass in the light-bino case (left),
and against the inelastic splitting in both cases (right). The bands around the predictions include theoretical and parametric uncertainties.}
\label{fig:mh}
\end{figure}
{\it Conclusion $-$} The combined ("bare" and dynamical) $\mu$ term allows a nearly pure higgsino DM  with  mass of about $1$~TeV and  neutral splitting. With each gaugino hierarchy, the splitting fixes the lighter gaugino mass. Matching  the supergravity boundary conditions and  the RG evolution through the EFT towers~(\ref{eq:tower}) yield
\begin{align}
&M_h=126.64\pm0.50\,({\rm th})\pm0.65\,(\alpha_s)\pm0.62\,(M_t)\ \mbox{GeV}
\qquad(\mbox{light bino}), \nonumber\\
&M_h=131.37\pm0.47\,({\rm th})\pm0.58\,(\alpha_s)\pm0.56\,(M_t)\ \mbox{GeV}
\qquad(\mbox{light wino}) ,
\label{eq:mh}
\end{align}
where the theoretical uncertainty combines the low-energy, high-scale matching and light-gaugino threshold contributions. The parametric uncertainties follow from $\alpha_s(M_Z)=0.1180\pm0.0009$ and $M_t=172.52\pm0.33$~GeV. When these uncertainties are added linearly, the light-bino prediction agrees with the measured Higgs boson  mass $125.13\pm0.11$~GeV~\cite{ParticleDataGroup:2026}, whereas the light-wino prediction is about 5 GeV above that. The wino triplet contributes  stronger than the bino to the running of the Higgs quartic coupling,  and  the former raises the predicted Higgs mass. Recent solar-capture bounds exclude the higgsino interpretation of the LZ event by requiring larger splittings but do not question higgsino DM itself ~\cite{Pospelov:2026solar}. As is shown in Fig.~\ref{fig:mh}, the light-bino case remains compatible with the measured Higgs mass within the quoted uncertainties for larger splittings up to $\delta=1$~MeV in the range considered. The light-bino case realizes higgsino DM in the same framework ~\cite{Frolovsky:2026akv} for inflation, spontaneous SUSY breaking and particle physics beyond the Standard Model. 

{\it Acknowledgements $-$} The work of DF was funded by the NWA ORC programme Emergence at All Scales. SVK was partially supported by Tokyo Metropolitan University and the World Premier International Research Center Initiative, MEXT, Japan. 



\bibliography{Bibliography}{}

@article{Frolovsky:2026akv,
  author = {Frolovsky, Daniel and Belyaev, Alexander and Ketov, Sergei V.},
  title = {{Higgs boson mass and thermal wino dark matter from Starobinsky supergravity with the MSSM}},
  eprint = {2607.07193},
  archivePrefix = {arXiv},
  primaryClass = {hep-ph},
  reportNumber = {IPMU26-0018},
  year = {2026}
}

@misc{LZ,
  author = {Akerib, D. S. and others},
  collaboration = {LUX-ZEPLIN},
  title = {{Search for dark matter particle interactions in an extended nuclear recoil energy window with the LUX-ZEPLIN (LZ) experiment}},
  howpublished = {\href{https://lz.lbl.gov/wp-content/uploads/sites/6/2026/08/LZ_Preprint_260901_Dark_Matter_EFT_Nuclear_Recoil_Search_at_Higher_Energies.pdf}{Collaboration preprint}},
  year = {2026}
}

@article{FT,
  author = {Freese, Katherine and Theodosopoulos, Dionysios P.},
  title = {{Higgsino Dark Matter Interpretation of the LUX-ZEPLIN 248 keV Nuclear-Recoil Event}},
  eprint = {2609.01583},
  archivePrefix = {arXiv},
  primaryClass = {hep-ph},
  year = {2026}
}

@article{Yin,
  author = {Yin, Wen},
  title = {{A PQ-Symmetric High-Scale SUSY Interpretation of the LZ High-Energy Recoil}},
  eprint = {2609.01892},
  archivePrefix = {arXiv},
  primaryClass = {hep-ph},
  year = {2026}
}

@article{DiMauro,
  author = {Di Mauro, Mattia},
  title = {{Dark Matter at the Kinematic Edge: Interpreting the 248 keV LZ Nuclear-Recoil Candidate}},
  eprint = {2609.02608},
  archivePrefix = {arXiv},
  primaryClass = {hep-ph},
  year = {2026}
}

@article{Brignole:1997wnc,
    author = "Brignole, A. and Ibanez, Luis E. and Munoz, C.",
    title = "{Soft supersymmetry breaking terms from supergravity and superstring models}",
    eprint = "hep-ph/9707209",
    archivePrefix = "arXiv",
    reportNumber = "CERN-TH-97-143, FTUAM-97-7, KAIST-TH-97-7",
    doi = "10.1142/9789812839657_0003",
    journal = "Adv. Ser. Direct. High Energy Phys.",
    volume = "18",
    pages = "125--148",
    year = "1998"
}

@article{Brignole:2010sax,
    author = "Brignole, A. and Ibanez, Luis E. and Munoz, C.",
    editor = "Kane, Gordon L.",
    title = "{Soft supersymmetry breaking terms from supergravity and superstring models}",
    doi = "10.1142/9789814307505_0004",
    journal = "Adv. Ser. Direct. High Energy Phys.",
    volume = "21",
    pages = "244--268",
    year = "2010"
}

@article{Safdi:2025sfs,
    author = {Safdi, Benjamin R. and Xu, Weishuang Linda},
    title = {{Wino and Real Minimal Dark Matter Excluded by Fermi Gamma-Ray Observations}},
    eprint = {2507.15934},
    archivePrefix = {arXiv},
    primaryClass = {hep-ph},
    year = {2025}
}

@incollection{Ketov:2025nkr,
    author = {Ketov, Sergei V.},
    title = {{On Legacy of Starobinsky Inflation}},
    booktitle = {Open Issues in Gravitation and Cosmology},
    publisher = {Springer},
    year = {2026},
    eprint = {2501.06451},
    archivePrefix = {arXiv},
    primaryClass = {gr-qc},
    isbn = {978-3-032-15702-7}
}

@article{Starobinsky:1980te,
	author = {Starobinsky, Alexei A.},
	doi = {10.1016/0370-2693(80)90670-X},
	issn = {0370-2693},
	journal = {Phys. Lett. B},
	number = {1},
	pages = {99 - 102},
	title = {A new type of isotropic cosmological models without singularity},
	volume = {91},
	year = {1980}}

@article{BICEP:2021xfz,
	archiveprefix = {arXiv},
	author = {Ade, P. A. R. and others},
	collaboration = {BICEP, Keck},
	doi = {10.1103/PhysRevLett.127.151301},
	eprint = {2110.00483},
	journal = {Phys. Rev. Lett.},
	number = {15},
	pages = {151301},
	primaryclass = {astro-ph.CO},
	title = {{Improved Constraints on Primordial Gravitational Waves using Planck, WMAP, and BICEP/Keck Observations through the 2018 Observing Season}},
	volume = {127},
	year = {2021}}

@article{Cecotti:1987sa,
	author = {Cecotti, S.},
	doi = {10.1016/0370-2693(87)90844-6},
	journal = {Phys. Lett.},
	pages = {86-92},
	reportnumber = {CERN-TH-4650-87},
	slaccitation = {%%CITATION = PHLTA,B190,86;%%},
	title = {{Higher derivative supergravity is equivalent to standard supergravity coupled to matter. 1.}},
	volume = {B190},
	year = {1987}}

@article{Gates:2009hu,
	archiveprefix = {arXiv},
	author = {Gates, Jr., S. James and Ketov, Sergei V.},
	doi = {10.1016/j.physletb.2009.03.005},
	eprint = {0901.2467},
	journal = {Phys. Lett. B},
	pages = {59--63},
	primaryclass = {hep-th},
	reportnumber = {UMDEPP-09-023},
	title = {{Superstring-inspired supergravity as the universal source of inflation and quintessence}},
	volume = {674},
	year = {2009}}

@article{Ketov:2010qz,
	archiveprefix = {arXiv},
	author = {Ketov, Sergei V. and Starobinsky, Alexei A.},
	doi = {10.1103/PhysRevD.83.063512},
	eprint = {1011.0240},
	journal = {Phys. Rev. D},
	pages = {063512},
	primaryclass = {hep-th},
	reportnumber = {RESCEU-24-10},
	title = {{Embedding $(R+R^{2})$-Inflation into Supergravity}},
	volume = {83},
	year = {2011}}

@article{Farakos:2013cqa,
	archiveprefix = {arXiv},
	author = {Farakos, F. and Kehagias, A. and Riotto, A.},
	doi = {10.1016/j.nuclphysb.2013.08.005},
	eprint = {1307.1137},
	journal = {Nucl. Phys. B},
	pages = {187--200},
	primaryclass = {hep-th},
	title = {{On the Starobinsky Model of Inflation from Supergravity}},
	volume = {876},
	year = {2013}}

@article{Ketov:2013dfa,
	archiveprefix = {arXiv},
	author = {Ketov, Sergei V. and Terada, Takahiro},
	doi = {10.1007/JHEP12(2013)040},
	eprint = {1309.7494},
	journal = {JHEP},
	pages = {040},
	primaryclass = {hep-th},
	reportnumber = {IPMU13-0185, UT-13-35},
	slaccitation = {%%CITATION = ARXIV:1309.7494;%%},
	title = {{Old-minimal supergravity models of inflation}},
	volume = {12},
	year = {2013}}

@article{Kehagias:2013mya,
    author = "Kehagias, Alex and Moradinezhad Dizgah, Azadeh and Riotto, Antonio",
    title = "{Remarks on the Starobinsky model of inflation and its descendants}",
    eprint = "1312.1155",
    archivePrefix = "arXiv",
    primaryClass = "hep-th",
    doi = "10.1103/PhysRevD.89.043527",
    journal = "Phys. Rev. D",
    volume = "89",
    number = "4",
    pages = "043527",
    year = "2014"
}

@article{Addazi:2017rkc,
	archiveprefix = {arXiv},
	author = {Addazi, Andrea and Ketov, Sergei V.},
	doi = {10.1088/1475-7516/2017/03/061},
	eprint = {1701.02450},
	journal = {JCAP},
	pages = {061},
	primaryclass = {hep-th},
	title = {{Energy conditions in Starobinsky supergravity}},
	volume = {03},
	year = {2017}}

@article{Ellis:2018zya,
    author = "Ellis, John and Nanopoulos, Dimitri V. and Olive, Keith A. and Verner, Sarunas",
    title = "{A general classification of Starobinsky-like inflationary avatars of SU(2,1)/SU(2) $\times$ U(1) no-scale supergravity}",
    eprint = "1812.02192",
    archivePrefix = "arXiv",
    primaryClass = "hep-th",
    reportNumber = "KCL-PH-TH-2018-69, KCL-PH-TH/2018-69, CERN-TH-2018-260, ACT-04-18, MI-TH-1813, UMN-TH-3806-18, FTPI-MINN-18-21, MI-TH-1813,
  UMN-TH-3806/18, FTPI-MINN-18/21",
    doi = "10.1007/JHEP03(2019)099",
    journal = "JHEP",
    volume = "03",
    pages = "099",
    year = "2019"
}

@article{Antoniadis:2024ypf,
    author = "Antoniadis, Ignatios and Nanopoulos, Dimitri V. and Olive, Keith A.",
    title = "{R$^{2}$-inflation derived from 4d strings, the role of the dilaton, and turning the Swampland into a mirage}",
    eprint = "2410.16541",
    archivePrefix = "arXiv",
    primaryClass = "hep-th",
    reportNumber = "UMN--TH--4402/24, FTPI--MINN--24/22, CERN-TH-2024-175",
    doi = "10.1007/JHEP06(2025)155",
    journal = "JHEP",
    volume = "06",
    pages = "155",
    year = "2025"
}

@article{Dalianis:2014aya,
	archiveprefix = {arXiv},
	author = {Dalianis, I. and Farakos, F. and Kehagias, A. and Riotto, A. and von Unge, R.},
	doi = {10.1007/JHEP01(2015)043},
	eprint = {1409.8299},
	journal = {JHEP},
	pages = {043},
	primaryclass = {hep-th},
	title = {{Supersymmetry Breaking and Inflation from Higher Curvature Supergravity}},
	volume = {01},
	year = {2015}}

@article{Dessert:2022evk,
    author = "Dessert, Christopher and Foster, Joshua W. and Park, Yujin and Safdi, Benjamin R. and Xu, Weishuang Linda",
    title = "{Higgsino Dark Matter Confronts 14~Years of Fermi {\ensuremath{\gamma}}-Ray Data}",
    eprint = "2207.10090",
    archivePrefix = "arXiv",
    primaryClass = "hep-ph",
    reportNumber = "MIT-CTP/5454",
    doi = "10.1103/PhysRevLett.130.201001",
    journal = "Phys. Rev. Lett.",
    volume = "130",
    number = "20",
    pages = "201001",
    year = "2023"
}

@article{Staub:2013tta,
    author = {Staub, Florian},
    title = {{SARAH 4: A tool for (not only SUSY) model builders}},
    eprint = {1309.7223},
    archivePrefix = {arXiv},
    primaryClass = {hep-ph},
    doi = {10.1016/j.cpc.2014.02.018},
    journal = {Comput. Phys. Commun.},
    volume = {185},
    pages = {1773--1790},
    year = {2014}
}

@article{Athron:2014yba,
    author = {Athron, Peter and Park, Jae-hyeon and St{\"o}ckinger, Dominik and Voigt, Alexander},
    title = {{FlexibleSUSY---A spectrum generator generator for supersymmetric models}},
    eprint = {1406.2319},
    archivePrefix = {arXiv},
    primaryClass = {hep-ph},
    doi = {10.1016/j.cpc.2014.12.020},
    journal = {Comput. Phys. Commun.},
    volume = {190},
    pages = {139--172},
    year = {2015}
}

@article{Athron:2017fvs,
    author = {Athron, Peter and Bach, Markus and Harries, Dylan and Kwasnitza, Thomas and Park, Jae-hyeon and St{\"o}ckinger, Dominik and Voigt, Alexander and Ziebell, Jobst},
    title = {{FlexibleSUSY 2.0: Extensions to investigate the phenomenology of SUSY and non-SUSY models}},
    eprint = {1710.03760},
    archivePrefix = {arXiv},
    primaryClass = {hep-ph},
    doi = {10.1016/j.cpc.2018.04.016},
    journal = {Comput. Phys. Commun.},
    volume = {230},
    pages = {145--217},
    year = {2018}
}

@article{Bagnaschi:2017xid,
    author = {Bagnaschi, Emanuele and Pardo Vega, Javier and Slavich, Pietro},
    title = {{Improved determination of the Higgs mass in the MSSM with heavy superpartners}},
    eprint = {1703.08166},
    archivePrefix = {arXiv},
    primaryClass = {hep-ph},
    doi = {10.1140/epjc/s10052-017-4885-7},
    journal = {Eur. Phys. J. C},
    volume = {77},
    pages = {334},
    year = {2017}
}

@article{FuentesMartin:2022matchete,
    author = {Fuentes-Mart{\'i}n, Javier and K{\"o}nig, Matthias and Pag{\`e}s, Julie and Thomsen, Anders Eller and Wilsch, Felix},
    title = {{A proof of concept for Matchete: an automated tool for matching effective theories}},
    eprint = {2212.04510},
    archivePrefix = {arXiv},
    primaryClass = {hep-ph},
    doi = {10.1140/epjc/s10052-023-11726-1},
    journal = {Eur. Phys. J. C},
    volume = {83},
    number = {7},
    pages = {662},
    year = {2023}
}

@article{Ibarra:2024weinberg,
    author = {Ibarra, Alejandro and Leister, Nicholas and Zhang, Di},
    title = {{Complete two-loop renormalization group equation of the Weinberg operator}},
    eprint = {2411.08011},
    archivePrefix = {arXiv},
    primaryClass = {hep-ph},
    doi = {10.1007/JHEP03(2025)214},
    journal = {JHEP},
    volume = {03},
    pages = {214},
    year = {2025}
}

@article{ParticleDataGroup:2026,
    author = {Takahashi, F. and others},
    collaboration = {Particle Data Group},
    title = {{Review of Particle Physics}},
    journal = {Int. J. Mod. Phys. A},
    volume = {41},
    pages = {2630011},
    year = {2026},
    doi = {10.1142/S0217751X26300115}
}

@article{Pospelov:2026solar,
  author = {Pospelov, Maxim and Ramani, Harikrishnan},
  title = {{Strong Constraints on Higgsino Dark Matter from Solar Capture}},
  eprint = {2609.02775},
  archivePrefix = {arXiv},
  primaryClass = {hep-ph},
  year = {2026}
}

@article{IceCube:2025fcu,
  author = {Abbasi, R. and others},
  collaboration = {IceCube},
  title = {{Search for High-Energy Neutrinos From the Sun Using Ten Years of IceCube Data}},
  eprint = {2507.08457},
  archivePrefix = {arXiv},
  primaryClass = {hep-ex},
  month = {7},
  year = {2025}
}
\bibliographystyle{utphys}

\end{document}